\documentclass[prd,superscriptaddress,amsfonts,amssymb,amsmath,showpacs,twocolumn]{revtex4-2}
\usepackage{bm}
\usepackage{amsfonts}
\usepackage{latexsym}
\usepackage[latin1]{inputenc}
\usepackage{graphicx}
\usepackage{amsmath}
\usepackage{palatino}
\usepackage{mathpazo}
\usepackage{textcomp}
\usepackage{float}
\usepackage{booktabs}
\usepackage{dcolumn}
\usepackage{booktabs}
\usepackage{multirow}
\usepackage{hyperref}
\hypersetup{colorlinks,citecolor=blue}
\usepackage{amsmath}
\usepackage{xcolor}
\usepackage{orcidlink}
\usepackage[caption=false]{subfig}
\usepackage{commath}
\def\jnl@style{\it}
\def\aaref@jnl#1{{\jnl@style#1}}

\def\aaref@jnl#1{{\jnl@style#1}}

\def\aj{\aaref@jnl{AJ}}                   
\def\apj{\aaref@jnl{ApJ}}                 
\def\apjl{\aaref@jnl{ApJ}}                
\def\apjs{\aaref@jnl{ApJS}}               
\def\apss{\aaref@jnl{Ap\&SS}}             
\def\aap{\aaref@jnl{A\&A}}                
\def\aapr{\aaref@jnl{A\&A~Rev.}}          
\def\aaps{\aaref@jnl{A\&AS}}              
\def\mnras{\aaref@jnl{Mon.~Not.~Roy.~Astron.~Soc.}}             
\def\prd{\aaref@jnl{Phys.~Rev.~D}}        
\def\prc{\aaref@jnl{Phys.~Rev.~C}}  
\def\prl{\aaref@jnl{Phys.~Rev.~Lett.}}    
\def\qjras{\aaref@jnl{QJRAS}}             
\def\skytel{\aaref@jnl{S\&T}}             
\def\ssr{\aaref@jnl{Space~Sci.~Rev.}}     
\def\zap{\aaref@jnl{ZAp}}                 
\def\nat{\aaref@jnl{Nature}}              
\def\aplett{\aaref@jnl{Astrophys.~Lett.}} 
\def\apspr{\aaref@jnl{Astrophys.~Space~Phys.~Res.}} 
\def\physrep{\aaref@jnl{Phys.~Rep.}}      
\def\physscr{\aaref@jnl{Phys.~Scr}}       
\def\commat{\aaref@jnl{Comm.~Math.~Phys.}}              
\def\science{\aaref@jnl{Science}}               
\def\cqg{\aaref@jnl{Classical Quant.~Grav.}}            
\def\jpcs{\aaref@jnl{JPCS}}                                     
\def\ijmpd{\aaref@jnl{Int.~J.~Mod.~Phys.~D}}                    
\def\grg{\aaref@jnl{Gen.~Relat.~Gravit.}}               
\def\rpp{\aaref@jnl{Rep.~Prog.~Phys.}}          
\def\npa{\aaref@jnl{Nucl.~Phys.~A}}        
\def\lrr{\aaref@jnl{Living Rev.~Rel.}}                   
\def\jcap{\aaref@jnl{J.~Cosmology Astropart.~Phys.}}    
\def\rmp{\aaref@jnl{Rev.~Mod.~Phys.}}   
\def\epjc{\aaref@jnl{Eur.~Phys.~J.~C}}

\allowdisplaybreaks[1]

\begin{document}

\color{black}       

\title{Cosmological constraints on dark energy in $f(Q)$ gravity: A parametrized perspective}

\author{A. Mussatayeva \orcidlink{0000-0000-0000-0000}}
\email[Email: ]{a.b.mussatayeva@gmail.com}
\affiliation{
Department of Physics and Chemistry, S. Seifullin Kazakh Agrotechnical University, Astana 010011, Kazakhstan}

\author{N. Myrzakulov\orcidlink{0000-0001-8691-9939}}
\email[Email: ]{nmyrzakulov@gmail.com}
\affiliation{L. N. Gumilyov Eurasian National University, Astana 010008,
Kazakhstan.}
\affiliation{Ratbay Myrzakulov Eurasian International Centre for Theoretical
Physics, Astana 010009, Kazakhstan.}

\author{M. Koussour\orcidlink{0000-0002-4188-0572}}
\email[Email: ]{pr.mouhssine@gmail.com}
\affiliation{Quantum Physics and Magnetism Team, LPMC, Faculty of Science Ben
M'sik,\\
Casablanca Hassan II University,
Morocco.}

\date{\today}

\begin{abstract}
In this paper, we focus on the parametrization of the effective equation of state (EoS) parameter within the framework of $f(Q)$ symmetric teleparallel gravity. Here, the gravitational action is represented by an arbitrary function of the non-metricity scalar $Q$. By utilizing a specific parametrization of the effective EoS parameter and a power-law model of $f(Q)$ theory, namely $f(Q)=\beta Q^{\left( m+1\right) }$ (where $\beta$ and $m$ are arbitrary constants), we derive the cosmological solution of the Hubble parameter $H(z)$. To constrain model parameters, we employ recent observational data, including the Observational Hubble parameter Data ($OHD$), Baryon Acoustic Oscillations data ($BAO$), and Type Ia supernovae data ($SNe$ Ia). The current constrained value of the deceleration parameter is found to be $q_{0}=-0.50^{+0.01}_{-0.01}$, indicating that the current Universe is accelerating. Furthermore, we examine the evolution of the density, EoS, and $Om(z)$ diagnostic parameters to deduce the accelerating nature of the Universe. Finally, we perform a stability analysis with linear perturbations to confirm the model's stability.
\end{abstract}

\maketitle

\section{Introduction}
\label{sec1}

In modern cosmology, the observational aspect is critical. The introduction of new tools in observation causes cosmologists to reassess the formulation of gravitational theories regularly. With the discovery of Hubble, Einstein was forced to remove the cosmological constant from his field equations in General Relativity Theory (GRT). The observation of Type Ia supernovae ($SNe$ Ia) in 1998 forced cosmologists to abandon the hypothesis of decelerating Universe expansion \cite{Riess,Perlmutter}. Since then, the Baryon Acoustic Oscillations ($BAO$) \cite{D.J.,W.J.}, Cosmic Microwave Background ($CMB$) \cite{R.R.,Z.Y.}, and Large Scale Structure ($LSS$) \cite{T.Koivisto,S.F.}, and many more measurements have provided evidence for the Universe's accelerated expansion. Thus, it is critical to include observable data while developing a theoretical cosmological model of the Universe. The accelerated expansion of the Universe is a key characteristic of modern cosmology. The Einstein field equations in GRT invariably result in a decelerating expansion of the Universe with the normal matter constituent. The accelerating expansion can be characterized by introducing a new constituent to the energy-momentum tensor part of the field equations or by making some changes to the geometrical part. Using these concepts, recent research has developed a variety of cosmological models of the Universe that explain the accelerating expansion. The notion of dark energy (DE) has recently gained prominence. DE is an exotic energy constituent with high negative pressure that explains numerous data and addresses several significant issues in modern cosmology. The second alternative is to suppose that GRT fails at large scales and that gravity may be explained via a more general action than the Einstein-Hilbert action.

In general, modified theories of gravity can be divided between models following the GRT structure with null torsion and non-metricity (such as the $f(R)$ and $f(R,T)$ theories \cite{Capo/2008,Nojiri/2007, Harko/2011,Momeni/2015}), models with torsion $T$ (the teleparallel
equivalent of GRT) \cite{Capo/2011,Nunes/2016}, and models with non-metricity $Q$ (the symmetric teleparallel equivalent of GRT) \cite{Q0,Q1}. Here, we will examine the $f(Q)$ theory, an extension of the symmetric teleparallel equivalent GRT in which gravity is due to the non-metricity scalar $Q$. In $f(Q)$ theory, the covariant divergence of the metric tensor $g_{\mu \nu }$ is non-zero, and this feature can be represented mathematically in terms of a new geometric variable known as non-metricity i.e. $Q_{\gamma \mu \nu }=\nabla _{\gamma }g_{\mu \nu }$, which geometrically represents the variation of the length of a vector in a parallel transport process. 

Recently, several intriguing cosmological and astrophysical consequences of $f(Q)$ gravity have been published, such as: The first cosmological solutions \cite{Q1,Q2}; Quantum cosmology \cite{Q3}; The coupling matter in $f(Q)$ gravity \cite{Q4}; Black hole solutions \cite{Q5}; General covariant symmetric teleparallel gravity \cite{Q6}; Evidence that non-metricity of $f(Q)$ gravity can challenge $\Lambda$CDM \cite{Q7}; Gravitational waves \cite{Q8,Q9,Q10}; The acceleration of the Universe and DE \cite{ko1,ko2,ko3,ko4,ko5}; Observational constraints \cite{OC1, OC2, OC3}.

Motivated by the previous discussion and studies on modified $f(Q)$ theory of gravity, in the present study, the accelerated expansion has been investigated using one specific parameterization of the total or effective equation of state (EoS) parameter $\omega_{eff}$ in the background of $f(Q)$ theory of gravity (Sec. \ref{sec3} explored the fundamental features of the specified $\omega_{eff}$). We have also considered the power-law form of $f(Q)=\beta Q^{\left( m+1\right) }$,
where $\beta$ and $m$ are arbitrary constants \cite{Q4}. The primary purpose of this research is to examine the nature of late-time cosmology's evolution. The observational constraints on model parameters are established by employing the Observational Hubble parameter data ($OHD$), $BAO$ data, and $SNe$ data. We then examined the evolution of the density parameter, the effective EoS parameter, and the deceleration parameter at the $1-\sigma$ and $2-\sigma$ confidence levels (CL) using the estimated values of model parameters. This work is structured as follows: in Sec. \ref{sec2}, we present a brief review of the $f(Q)$ gravity. In Sec. \ref{sec3}, we write the cosmological solution of the Hubble parameter by using a specific parameterization of the effective EoS parameter and a power-law model of $f(Q)$ theory. In Sec. \ref{sec4}, we calculate the values of the model parameters using the combined $OHD+BAO+SNe$ data. Moreover, we describe the behavior of several parameters such as the density, EoS, and deceleration parameters. In Sec. \ref{sec5}, we examine the $Om(z)$ diagnostic parameter history of our $f(Q)$ model to see if the assumed model recognizes the DE behavior, and then we do a linear perturbation analysis. Finally, in Sec. \ref{sec6}, we summarize our findings.

\section{A brief review of $f(Q)$ gravity}
\label{sec2}

In general, in the presence of matter components, the action for a $f(Q)$
gravity model is written as \cite{Q0,Q1}, 
\begin{equation}
S=\int \sqrt{-g}d^{4}x\left[ \frac{f(Q)}{2\kappa ^{2}}+L_{m}\right] ,
\label{1}
\end{equation}%
where $g$ is the determinant of the metric tensor $g^{\mu \nu }$, i.e. $%
g=det(g_{\mu \nu })$, $\kappa ^{2}=8\pi G=1/M_{p}^{2}$, $G$ is the Newtonian
constant, while $M_{p}$ is the reduced Planck mass. $L_{m}$ denotes the
Lagrangian density of the matter components. For the time being, the term $%
f(Q)$ is an arbitrary function of the non-metricity scalar $Q$.

The tensor of non-metricity and its traces are given by 
\begin{equation}
Q_{\gamma \mu \nu }=\nabla _{\gamma }g_{\mu \nu },  \label{4}
\end{equation}%
\begin{equation}
Q_{\beta }=g^{\mu \nu }Q_{\beta \mu \nu }\qquad \widetilde{Q}_{\beta
}=g^{\mu \nu }Q_{\mu \beta \nu }.  \label{5}
\end{equation}

Furthermore, as a function of the non-metricity tensor, the superpotential
(or the non-metricity conjugate) can be expressed as, 
\begin{equation}
P_{\,\,\,\mu \nu }^{\beta }=-\frac{1}{2}L_{\,\,\,\mu \nu }^{\beta }+\frac{1}{%
4}(Q^{\beta }-\widetilde{Q}^{\beta })g_{\mu \nu }-\frac{1}{4}\delta _{(\mu
}^{\beta }Q_{\nu )}.  \label{6}
\end{equation}%
where ${L^{\beta }}_{\mu \nu }$ is the disformation tensor,
\begin{equation}
{L^{\beta }}_{\mu \nu }\equiv \frac{1}{2}g^{\beta \sigma }\left( Q_{\nu
\mu \sigma }+Q_{\mu \nu \sigma }-Q_{\beta \mu \nu }\right) .
\end{equation}

Hence, the non-metricity scalar is expressed as, 
\begin{equation}
Q=-Q_{\beta \mu \nu }P^{\beta \mu \nu }\,.  \label{7}
\end{equation}

Using the variation of action in Eq. (\ref{1}) with respect to the metric tensor $%
g^{\mu \nu }$, one can obtain the field equations, 
\begin{multline}
\label{9}
\frac{2}{\sqrt{-g}}\nabla_{\beta}\left(f_{Q}\sqrt{-g}\,P^{\beta}_{\,\,\mu\nu}\right)+\frac{1}{2}f\,g_{\mu\nu}+\\
f_{Q}\left(P_{\mu\beta\lambda}Q_{\nu}^{\,\,\,\beta\lambda}-2Q_{\beta\lambda\mu}P^{\beta\lambda}_{\,\,\,\,\,\,\nu}\right)=- T_{\mu \nu},
\end{multline}
where $f_{Q}=\dfrac{df}{dQ}$. Moreover, $T_{\mu \nu }$ is the
energy-momentum tensor of the cosmic fluid, which is considered to be a
perfect fluid, i.e. $T_{\mu \nu }=(\rho +p)u_{\mu }u_{\nu }+pg_{\mu \nu }$,
where $u^{\mu }=(1,0,0,0)$ represents the 4-velocity vector components that form the fluid. $\rho $ and $p$ represent the total energy density and total pressure of any perfect fluid of matter and DE, respectively.

In the context of a flat FLRW space-time, the modified Friedmann
equations 
\begin{equation}
ds^{2}=-dt^{2}+a^{2}(t)[dx^{2}+dy^{2}+dz^{2}],  \label{3a}
\end{equation}%
where $a(t)$ is the scale factor of the Universe are given by \cite{Q4}%
\begin{equation}
3H^{2}=\frac{1}{2f_{Q}}\left( -\rho +\frac{f}{2}\right) ,  \label{F1}
\end{equation}%
\begin{equation}
\dot{H}+3H^{2}+\frac{\dot{f}_{Q}}{f_{Q}}H=\frac{1}{2f_{Q}}\left( p+\frac{f}{2%
}\right) ,  \label{F2}
\end{equation}
where $Q=6H^{2}$, and $H$ denotes the Hubble parameter, which estimates the
rate of expansion of the Universe. It is interesting to note that the
standard Friedmann equations of GR can be found if the function $f(Q)=-Q$ is
considered, i.e. $3H^{2}=\rho $ and $2\dot{H}+3H^{2}=-p$.

In our study, we consider a simplified cosmological scenario where the universe is composed of two main components: matter and DE. The matter is assumed to be fluid without pressure ($p_{m}=0$), while DE is considered to possess negative pressure, which is responsible for driving the observed cosmic acceleration. For this reason, we assume that $\rho=\rho_{m}+\rho_{DE}$ and $p=p_{DE}$. In addition, the equation of state (EoS) parameter is a quantity used in cosmology to
explain the properties of DE. The effective or total EoS parameter is defined as the ratio of the total pressure to the total energy density. In the context of our study, it takes into account contributions from various cosmic components, including DE and matter. Therefore, the effective EoS parameter, denoted as $\omega_{eff}$, is given by
\begin{equation}
\omega_{eff} =\frac{p}{\rho }=-1+\frac{\left( \overset{.}{H}+\frac{\overset{.}{%
f_{Q}}}{f_{Q}}H\right) \left( 2f_{Q}\right) }{\left( \frac{f}{2}%
-6H^{2}f_{Q}\right) }.  \label{3l}
\end{equation}

The above dot symbolizes the differentiation with regard to cosmic time $t$. Furthermore, the EoS parameter which combines the energy density and pressure of the DE component is,
\begin{equation}
    \omega_{DE}=\frac{p_{DE}}{\rho_{DE}}
\end{equation}

Now, in order to derive the matter conservation equation, we can be taking the trace of the field equation,
\begin{equation}
\overset{.}{\rho }_{m}+3\rho _{m}H=0,  \label{cm}
\end{equation}

By solving Eq. (\ref{cm}), we are able to derive the solution for the energy density of the matter $\rho_{m}$ as,%
\begin{equation}
\rho _{m}=\rho _{m0}a^{-3},
\end{equation}%
where $\rho_{m0}$ is the present value of the energy density of the matter.

\section{Late-time cosmological evolution via a specific type of EoS
parameter}
\label{sec3}

This section examines the Universe's evolution at late times using a
specific type of EoS parameter. However, the equations obtained from this
analysis are complex and require numerical solutions. To simplify the
implementation of such solutions, a change of variable is performed, where
the red-shift, $z$, is used as the dynamical variable instead of the cosmic
time $t$. One starting point that we can rely on is that $z=\frac{a_{0}}{%
a\left( t\right) }-1$, where $a_{0}$ is the present time of the scale factor. For simplicity, the scale factor is set to $1$ currently. It is not directly observable, but we can observe the ratio of the scale factor at different times to its value at the present time. The following
relationship may therefore be deduced: $\frac{d}{dt}=-H\left( z\right)
\left( 1+z\right) \frac{d}{dz}$. Thus, it is clear that.%
\begin{equation}
\overset{.}{H}=-H\left( z\right) \left( 1+z\right) H^{^{\prime }}\left(
z\right) ,
\end{equation}%
where, the symbol 'prime' represents differentiation with respect to the
red-shift variable, denoted by '$z$'.

In this context, it is evident that we can utilize only Eqs. (\ref{F1}) and (\ref{F2}) for
our analysis. However, rather than solving the ensuing equation for $H(z)$,
we can introduce an effective form of the EoS parameter, which is defined as
follows: $\omega _{eff}=-1+\frac{A}{A+B\left( 1+z\right) ^{-3}}$, where $A$
and $B$ are arbitrary constants. The reason behind selecting this particular
parametrization for $\omega _{eff}$ is that at high red-shift values $z\gg 1$
(early stages of cosmological evolution), $\omega _{eff}$ is nearly zero,
indicating the behavior of the EoS parameter for a pressureless fluid, such
as ordinary matter. As we move towards the present epoch ($z=0$), $\omega
_{eff}$ decreases gradually to negative values, leading to negative pressure
and an effective EoS value $\omega _{eff}=-\frac{B}{A+B}$. In this case, the
functional form of $\omega _{eff}$ is dependent on the specific values of $A$
and $B$. As a result, the form of $\omega _{eff}$ can effortlessly
incorporate the phases of cosmic evolution, including the early
matter-dominated era and the late-time DE-dominated era. The specific form mentioned, introduced in Ref. \cite{P0}, exhibits phantom-like behavior in the present epoch. Due to the presence of a large number of free parameters in the effective EoS parameter, we adopt a specific approach for the observational analysis. In order to constrain the model and facilitate the analysis, we fix the value of $n$ to be $3$. In literature, various
parametrization models of EoS for DE have been proposed and fitted to
observational data. Ref. \cite{P1} proposed an one-parameter family of EoS
DE model. Two-parameters family of EoS DE parametrizations, especially the
Chevallier-Polarski-Linder parametrization \cite{P2, P3}, the Linear
parametrization \cite{P3, P4, P5, P6}, the Logarithmic parametrization \cite%
{P7}, the Jassal-Bagla-Padmanabhan parametrization \cite{P8}, and the
Barboza-Alcaniz parametrization \cite{P9}, were also explored. Further, in 
\cite{P10, P11, P12} three and four parameters family of EoS DE
parametrizations are examined.

In this section, we will look at a specific cosmological model in $f(Q)$
gravity theory. We also look at how geometrical and physical cosmological
parameters such as energy density, pressure, and deceleration behave under $%
f(Q) $ gravity. In this study, we investigate the scenario where the
function $f(Q)$ can be expressed as, $f(Q)=\beta Q^{\left( m+1\right) }$,
where $\beta$ and $m$ are arbitrary constants \cite{Q4,PW1,PW2}. For the function $f(Q)$ we
obtain the expression $f_{Q}=\beta \left( 1+m\right) Q^{m}$ and $%
f_{QQ}=\beta \left( 1+m\right) mQ^{m-1}$. By putting the above expressions
for $f$, $f_{Q}$, and $f_{QQ}$ into Eqs. \eqref{F1} and \eqref{F2} we can
derive the energy density and pressure as, 
\begin{equation}
\rho =\beta \left( -2^{m}\right) 3^{m+1}(2m+1)H^{2(m+1)},
\label{rhoM}
\end{equation}%
and%
\begin{equation}
p=\beta 6^{m}(2m+1)H^{2m}\left( 2\dot{H}(m+1)+3H^{2}\right)
.
\end{equation}

Now by using Eq. (\ref{3l}), we obtain the EoS parameter in terms of Hubble
parameter and its derivative as, 
\begin{equation}
\omega_{eff} =-1+\frac{2\left( m+1\right) }{3}\frac{\left( 1+z\right)
H^{\prime }\left( z\right) }{H\left( z\right) }.
\label{17}
\end{equation}

By using Eq. (\ref{17}) and the presumed ansatz of $\omega _{eff}$, the
evolution equation of the Hubble function takes the form 
\begin{equation}
\frac{dH\left( z\right) }{dz}=\frac{3A\left( 1+z\right) ^{2}}{2\left(
m+1\right) \left( A\left( 1+z\right) ^{3}+B\right) }H\left( z\right), 
\end{equation}%
which yields the following solution 
\begin{equation}
H\left( z\right) =H_{0}\left[ \frac{A(z+1)^{3}+B}{A+B}\right] ^{\frac{1}{2m+2%
}},  \label{H1}
\end{equation}%
where $H_{0}$ describes the present value (i.e. at $z=0$) of the Hubble
parameter. In particular, for the scenario $m=0$ with $\beta =-1$, the
solution reduces to $f\left( Q\right) =-Q$. In other words, it is directly
related to the $\Lambda $CDM model. As a result, the equation for Hubble
parameter $H\left( z\right) $ is reduced to $H\left( z\right) =H_{0}\left[
\Omega _{m}^{0}\left( 1+z\right) ^{3}+\Omega _{\Lambda }^{0}\right] ^{\frac{1%
}{2}}$, where $\Omega _{m}^{0}=\frac{A}{A+B}$ and $\Omega _{\Lambda
}^{0}=\left( 1-\Omega _{m}^{0}\right) =\frac{B}{A+B}$ are the present
density parameters for matter and the cosmological constant, respectively.
As a result, the model parameter $m$ is an excellent indicator of the
present model's deviation from the $\Lambda $CDM model due to the addition
of non-metricity terms.

The deceleration parameter $q$ is one of the cosmological parameters that is
important in describing the status of our Universe's expansion. If the value
of the deceleration parameter is strictly less than zero, the cosmos
accelerates; when it is non-negative, the cosmos decelerates. The
deceleration parameter $q$ is defined as $q\left( z\right) =-\frac{\overset{%
..}{a}}{aH^{2}}=-1+\frac{\left( 1+z\right) }{H\left( z\right) }\frac{%
dH\left( z\right) }{dz}$. In this scenario, the expression of the
deceleration parameter is%
\begin{equation}
q\left( z\right) =-1+\frac{3A(1+z)^{3}}{2(m+1)\left( A(1+z)^{3}+B\right) }.
\label{qz1}
\end{equation}

The behavior and important cosmological features of the model represented in
Eq. (\ref{H1}) are entirely reliant on the model parameters ($A$, $B$, and $m
$). In the next part, we use current observational data to study the
behavior of the cosmological parameters to constrain the model parameters ($A
$, $B$, and $m$).

\section{Method of data fitting}
\label{sec4}

In our research, we took into account the most current and relevant
observational findings:

\begin{itemize}
\item \textbf{Observational Hubble parameter Data ($OHD$)}: We examine $H(z)$ data
points calculated by employing differential galaxy ages as a function of
red-shift $z$ and line-of-sight $BAO$ data. \cite%
{Yu/2018,Moresco/2015,Sharov/2018}.

\item \textbf{Baryon Acoustic Oscillation ($BAO$)}: We additionally take into
account the $BAO$ data from the SDSS-MGS, Wiggle Z, and 6dFGS projects \cite%
{Blake/2011,Percival/2010,Giostri/2012}.

\item \textbf{Type-Ia Supernova measurement ($SNe$ Ia)}: We examine the
Pantheon sample of 1048 SNe Ia luminosity distance values from the
Pan-STARSS1 (PS1) Medium Deep Survey, the Low-z, SDSS, SNLS, and HST
missions \cite{Scolnic/2018,Chang/2019}.
\end{itemize}

In addition, for likelihood minimization, we employ the MCMC (Markov Chain Monte
Carlo) sample from the Python package \textit{emcee} \cite{Mackey/2013}%
, which is commonly used in astrophysics and cosmology to investigate the
parameter space $\theta_{s}=(A, B, m)$. To do this, we are now focusing on
three data: $OHD$, $BAO$, and $SNe$ Ia data. We evaluate the priors on the
parameters $-10.0<A<10.0$, $-10.0<B<10.0$, and $-10.0<m<10.0$. To find out the
outcomes of our MCMC study, we employed 100 walkers and 1000 steps. The discussion about the observational data has also been presented in a very similar fashion in Ref. 2, shedding further light on the significance of these findings. In the following subsections of our manuscript, we provide further detailed discussions on the observational data used, as well as the statistical analyses employed. We aim to present a comprehensive and transparent description of our methodology, emphasizing the novelty and contributions of our work while acknowledging the commonalities with existing literature.

\subsection{$OHD$}

We utilize a commonly popular compilation with an updated set of 57 data
points. In this collection of 57 Hubble data points, 31 were measured using
the method of differential age (DA), while the remaining 26 were measured
using $BAO$ and other methods in the red-shift range provided as $0.07\leq
z\leq2.42$, allowing us to determine the expansion rate of the Universe at
red-shift $z$. Hence, the Hubble parameter $H(z)$ as a function of red-shift
can be written as 
\begin{equation}
H(z)= -\frac{1}{1+z} \frac{dz}{dt}.
\end{equation}

To calculate the mean values of the model parameters  $A$, $B$, and $m$, we
used the chi-square function ($\chi^{2}$) for $OHD$ as, 
\begin{equation}
\chi^{2}_{OHD} = \sum_{i=1}^{57} \frac{\left[H(\theta_{s}, z_{i})-
H_{obs}(z_{i})\right]^2}{\sigma(z_{i})^2},
\end{equation}
where $H(z_{i})$ denotes the theoretical value for a specific model at
different red-shifts $z_{i}$, and $H_{obs}(z_{i})$ denotes the observational
value, $\sigma(z_{i})$ denotes the observational error.

\subsection{$BAO$}

We employ a compilation of SDSS, 6dFGS, and Wiggle Z surveys at various
red-shifts for $BAO$ data. This paper incorporates $BAO$ data as well as the
cosmology listed below, 
\begin{equation}
d_{A}(z)=c \int_{0}^{z} \frac{dy}{H(y)},
\end{equation}
\begin{equation}
D_{v}(z)=\left[ \frac{d_{A}^2 (z) c z }{H(z)} \right]^{1/3},
\end{equation}
where $d_{A}(z)$ represents the comoving angular diameter distance, and $%
D_{v}$ represents the dilation scale. Moreover, the chi-square function ($%
\chi^{2}$) for $BAO$ is given by 
\begin{equation}
\chi^{2}_{BAO} = X^{T} C_{BAO}^{-1} X.
\end{equation}
Here, X depends on the considered survey and $C_{BAO}^{-1}$ represents the
inverse covariance matrix \cite{Giostri/2012}.

\subsection{$SNe$}

To obtain the best values using SNe Ia, we begin with the measured distance
modulus $\mu_{obs}$ produced from SNe Ia detections and compare it to the
theoretical value $\mu_{th}$. The Pantheon sample, a recent SNe Ia dataset
containing 1048 points of distance modulus $\mu_{obs}$ at various red-shifts
in the range $0.01<z<2.26$, is taken into consideration in this work. The
distance modulus of each SNe can be calculated using the following
equations: 
\begin{equation}
\mu_{th}(z)=5 log_{10} \frac{d_{l}(z)}{Mpc}+25,
\end{equation}
\begin{equation}
d_{l}(z)=c(1+z) \int_{0}^{z} \frac{dy}{H(y,\theta)}.
\end{equation}
where $c$ is the speed of light. The distance modulus can be calculated
using the relationship, 
\begin{equation}
\mu= m_{B}-M_{B}+\alpha x_{1} - \beta\, c + \Delta_{M} + \Delta_{B},
\end{equation}
where $m_{B}$ is the measured peak magnitude at the B-band maximum, and $%
M_{B}$ is the absolute magnitude. The parameters $c$, $\alpha$, $\beta$, and 
$x_{1}$, respectively, correspond to the color at the brightness point, the
luminosity stretch-color relation, and the light color shape. Moreover, $%
\Delta_{M}$ and $\Delta_{B}$ are distance adjustments based on the host
galaxy's mass and simulation-based anticipated biases. The nuisance
parameters in the above equation were obtained using a novel method known as
BEAMS with Bias Corrections (BBC) \cite{Kessler/2017}. As a result, the
measured distance modulus is equal to the difference between the apparent
magnitude $m_{B}$ and the absolute magnitude $M_{B}$ i.e., $\mu = m_{B}-M_{B}
$. For the Pantheon data, the $\chi^{2}$ function is assumed to be,

\begin{equation}
\chi^{2}_{SNe} =\sum_{i,j=1} ^{1048} \Delta \mu_{i} \left(
C_{SNe}^{-1}\right)_{ij} \Delta \mu_{j}
\end{equation}
where $\Delta \mu_{i}= \mu_{th}-\mu_{obs}$ and $C_{SNe}$ represents the
covariance matrix.

\subsection{$OHD+BAO+SNe$}

Now, the $\chi^{2}$ function for the $OHD+BAO+SNe$ data is assumed to be, 

\begin{equation}
\chi^{2}_{total}=\chi^{2}_{OHD}+\chi^{2}_{BAO}+\chi^{2}_{SNe},
\end{equation}

By using the aforementioned combined $OHD+BAO+SNe$ data, we obtained the
best-fit values of the model parameters $A$, $B$, and $m$, as shown in Fig. %
\ref{H+SN+BAO1} with $1-\sigma$ and $2-\sigma$ likelihood contours. The
best-fit values obtained are $A=0.342^{+0.022}_{-0.022}$, $%
B=0.677^{+0.025}_{-0.025}$, and $m=0.013^{+0.021}_{-0.021}$. For $m=0$,
Fig. \ref{H+SN+BAO2} shows the results of $1-\sigma$ and $2-\sigma$
likelihood contours with the best-fit values of model parameters are $%
A=0.3353\pm 0.0010$, and $B=0.6837 \pm0.0019$.
Figs. \ref{ErrorHubble} and \ref{ErrorSNe} also show the error bars for $H(z)
$ and $\mu(z)$ using $H_{0}=(67.4\pm0.5)$ $Km/s/Mpc$  
\cite{Planck2020}. The figures also show a comparison of our model to the commonly
used $\Lambda$CDM model in cosmology i.e. $H\left( z\right) =H_{0}\sqrt{%
\Omega _{m}^{0}\left( 1+z\right) ^{3}+\Omega _{\Lambda}^{0}}$ (we have
considered $\Omega _{m}^{0}=0.315\pm 0.007$) \cite{Planck2020}. As shown in
the figures, our model matches the observed data nicely.

\begin{widetext}

\begin{figure}[h]
\centerline{\includegraphics[scale=0.8]{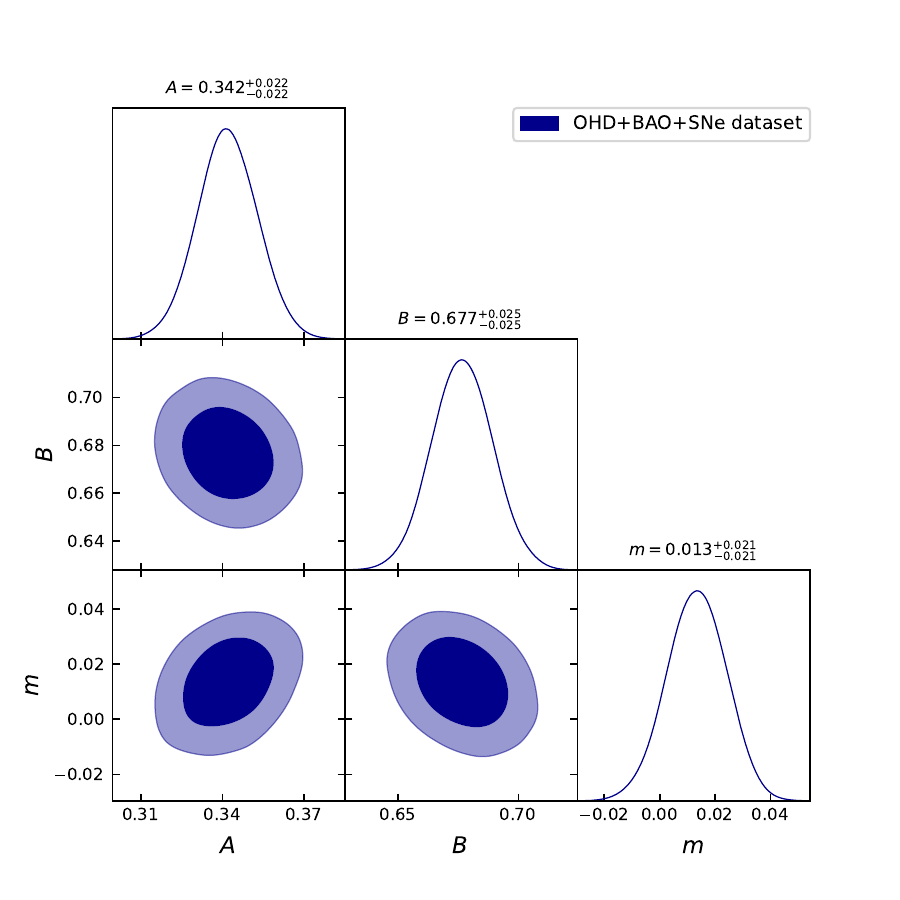}}
\caption{The $1-\sigma$ and $2-\sigma$ confidence curves for the model parameters $A$, $B$, and $m$ with combined $OHD+BAO+SNe$ data.}
\label{H+SN+BAO1}
\end{figure}

\begin{figure}[H]
\centerline{\includegraphics[scale=0.8]{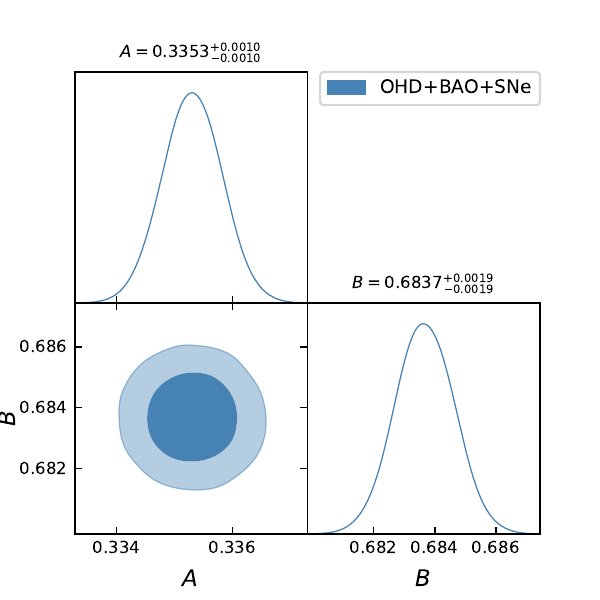}}
\caption{The $1-\sigma$ and $2-\sigma$ confidence curves for the model parameters $A$, and $B$ with combined $OHD+BAO+SNe$ data for $m=0$.}
\label{H+SN+BAO2}
\end{figure}

\begin{figure}[H]
\centerline{\includegraphics[scale=0.60]{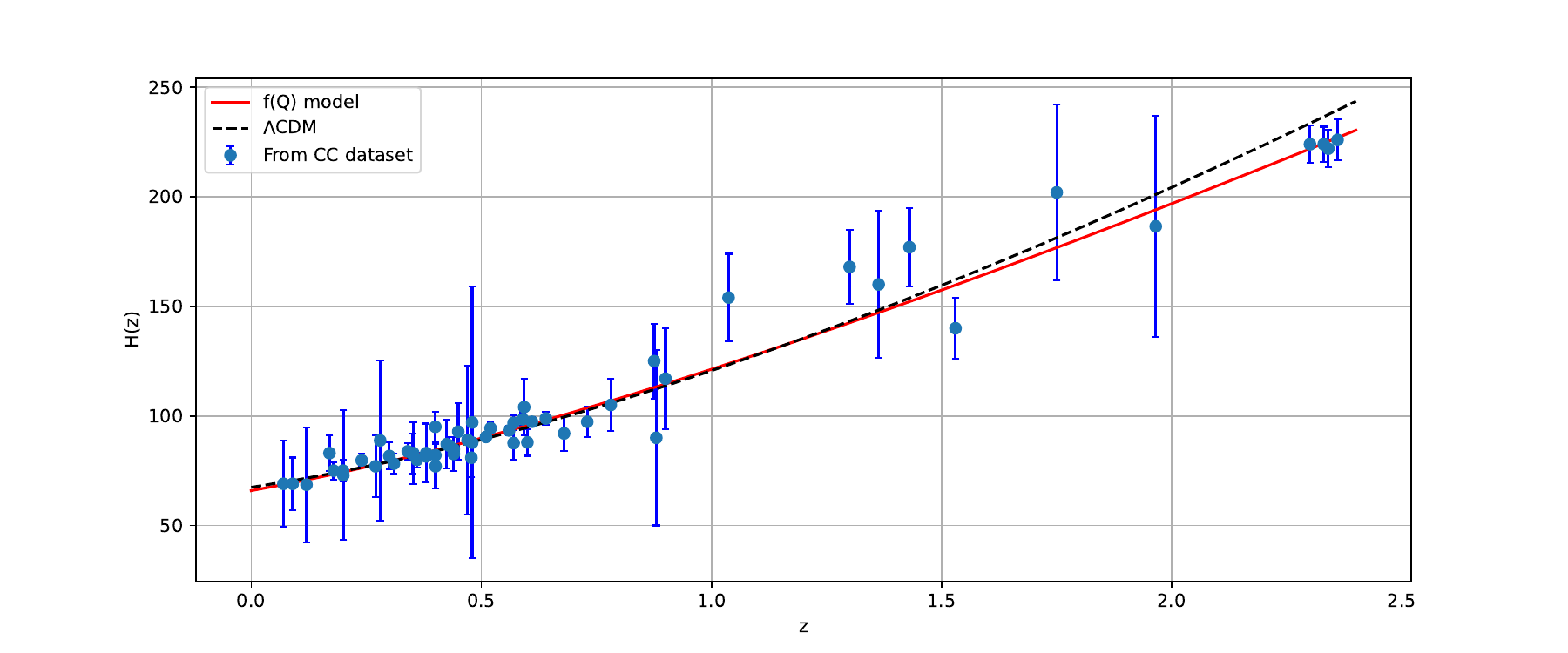}}
\caption{The evolution of Hubble parameter $H(z)$ with regard to red-shift $z$. The black dashed line represents the $\Lambda$CDM model, the red line is our model's curve, and the blue dots show error bars.}
\label{ErrorHubble}
\end{figure}

\begin{figure}[H]
\centerline{\includegraphics[scale=0.60]{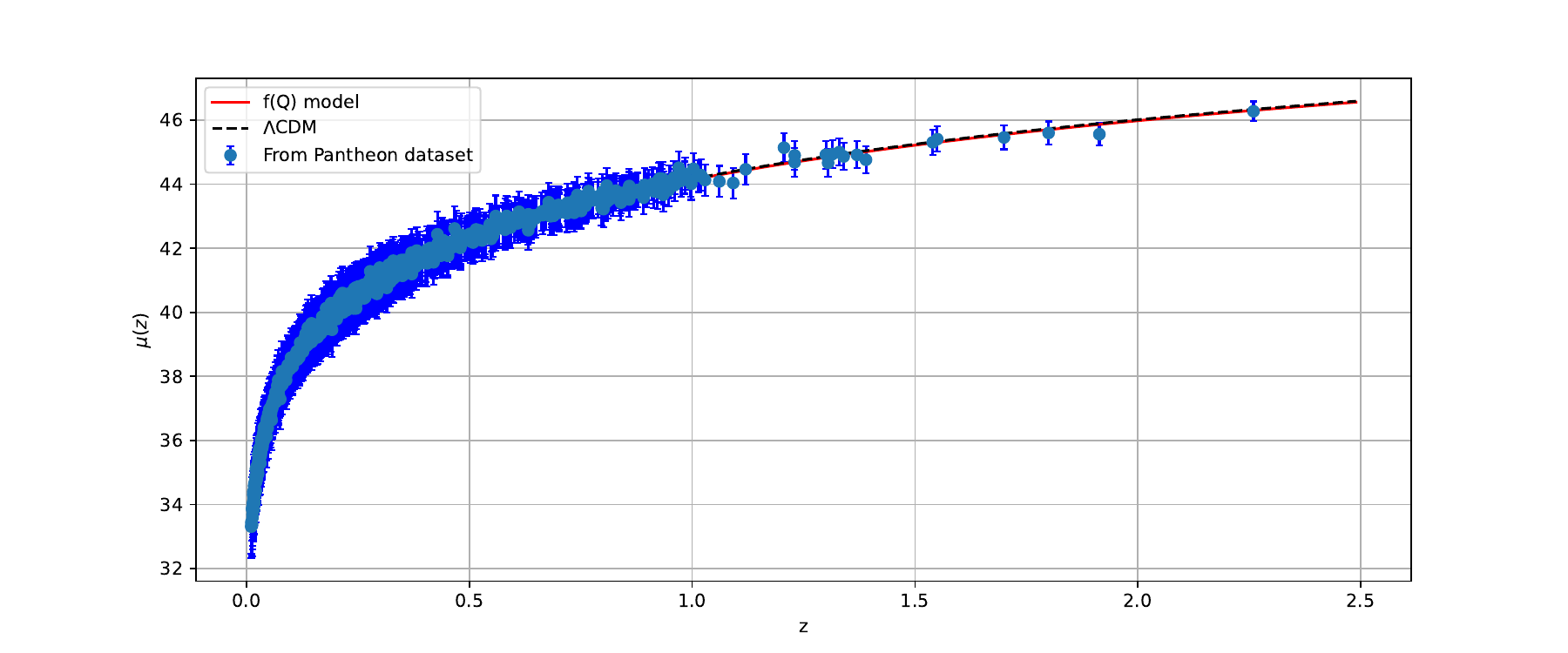}}
\caption{The evolution of distance modulus $\mu(z)$ with regard to red-shift $z$. The black dashed line represents the $\Lambda$CDM model, the red line is our model's curve, and the blue dots show error bars.}
\label{ErrorSNe}
\end{figure}
\end{widetext}

We will now discuss the cosmological consequences of the obtained
observational constraints. Using the obtained mean values of the model
parameters $A$, $B$, and $m$ constrained by the combined $OHD+BAO+SNe$ data,
we investigate the behavior of the density, the EoS, and the deceleration parameters.

\begin{figure}[h]
\centerline{\includegraphics[scale=0.70]{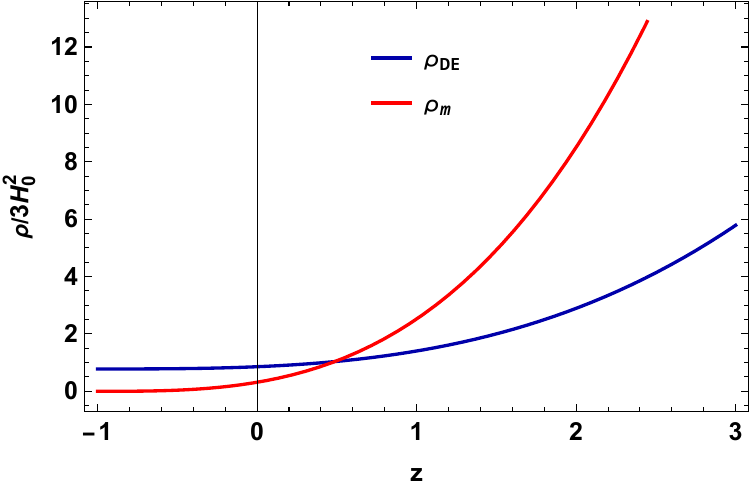}}
\caption{Evolution of the density parameter for matter and DE from the study of the combined $OHD+BAO+SNe$ data for the best fitting values of $A$, $B$, and $m$.}
\label{rho}
\end{figure}

\begin{figure}[h]
\centerline{\includegraphics[scale=0.70]{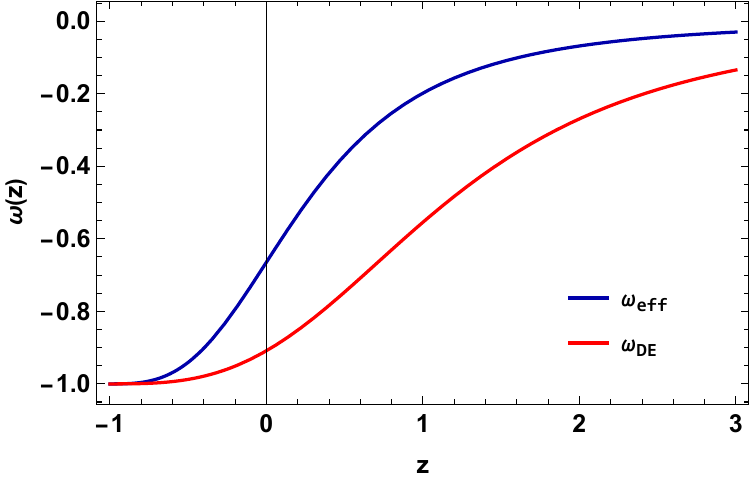}}
\caption{Evolution of the EoS parameter from the study of the
combined $OHD+BAO+SNe$ data for the best fitting values of $A$, and $B$.}
\label{Omega}
\end{figure}

\begin{figure}[h]
\centerline{\includegraphics[scale=0.70]{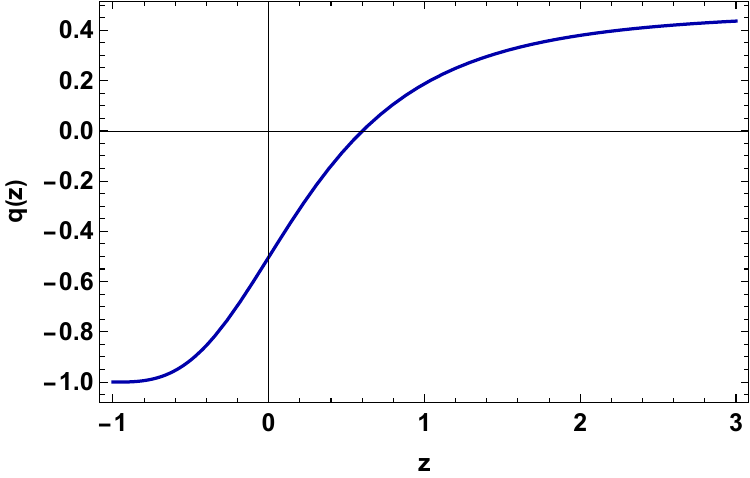}}
\caption{Evolution of the deceleration parameter from the study of the
combined $OHD+BAO+SNe$ data for the best fitting values of $A$, $B$, and $m$.}
\label{q}
\end{figure}

In Figs. \ref{rho}, \ref{Omega}, and \ref{q}, we presented the density
parameter, EoS parameter, and deceleration parameter as a function
of red-shift for the combined $OHD+BAO+SNe$ data. From Fig. \ref{rho}, it can be observed that as the universe expands, both the matter density parameter and the DE density parameter exhibit a decrease. In the late stages, the matter density approaches zero, while the DE density converges towards a small value. In addition, the densities parameter behaves positively for model parameter values constrained by the combined $OHD+BAO+SNe$ data.

As mentioned in Sec. \ref{sec2}, the EoS parameter is a vital cosmological
parameter for understanding the nature of the Universe and its history
through time, and it is defined as $\omega =\frac{p}{\rho }$, where $p$ is
the pressure and $\rho$ is the energy density. The value of the EoS
parameter governs how the fluid behaves and how it affects the expansion of
the Universe. For example, if $\omega = 0$, the fluid is referred to as
non-relativistic matter and behaves like dust. However, if $\omega = 1/3$,
the fluid is referred to as relativistic matter and behaves like radiation.
If $\omega <-1/3$, the fluid is considered to have negative pressure and is
responsible for the Universe's accelerated expansion, a phenomenon
associated with DE, which includes the quintessence $(-1< \omega < -1/3)$
era, cosmological constant $(\omega=-1)$, and phantom era $(\omega <-1)$.
The existing observational constraints imply that the EoS parameter of the
Universe's dominating component (DE), is extremely near to -1. In other
terms, the pressure of DE is negative and nearly constant, fueling the
Universe's accelerated expansion. Recent investigations of the $CMB$
radiation, the $LSS$ of the Universe, the luminosity-distance relation of $SNe$
Ia, and others, have given compelling evidence for the existence of DE and
its dominating role in the Universe's expansion. The most recent
measurements of the EoS parameter from these data produce a value of $%
\omega_{0}= -1.03 \pm 0.03$ \cite{Planck2020}, which is compatible with the
cosmological constant.

In this paper, we focus on the analysis of an effective EoS parameter using three model parameters: $A$, $B$, and $m$. The behavior of the EoS parameter is depicted
in Fig. \ref{Omega} for constrained values of $A$, $B$, and $m$ from the
combined $OHD+BAO+SNe$ data. From the analysis conducted, it is apparent that both the evolving EoS parameter for the DE and the effective EoS parameter demonstrate quintessence-like behavior. This observation highlights the resemblance to the typical characteristics associated with quintessence, shedding light on the intriguing nature of the DE component under investigation.
The present value (i.e. at $z=0$) of the EoS parameter for DE is $\omega_{0}=-0.91 \pm 0.08$ \cite{Hernandez,Gruber}, indicating an
accelerating phase.

In addition, as shown in Fig. \ref{q}, we analyzed the behavior of the
deceleration parameter for constrained values of $A$, $B$, and $m$ from the
combined $OHD+BAO+SNe$ data. The sign of the deceleration parameter $(q)$
indicates whether the model is accelerating or decelerating. If $q > 0$, the
model decelerates, if $q = 0$, it expands at a steady rate, and if $-1 < q <
0$, it expands at an accelerating rate. With $q = -1$, the Universe shows
exponential growth or De-Sitter expansion and super-exponential expansion
for $q < -1$. In Eq. (\ref{qz1}), we have obtained the deceleration
parameter for our model. According to Fig. \ref{q}, the model transitions
from a decelerated stage to an accelerated stage. It can also be seen that
our model initially decelerates $(q > 0)$ and then approaches exponential
expansion in late times $(q = -1)$. In the figure, we also compare our model to the commonly accepted $\Lambda$CDM model in cosmology. According to the constrained values of model
parameters $A$, $B$, and $m$ from the combined $OHD+BAO+SNe$ data, the present
value of the transition red-shift is $z_{tr}=0.60 \pm 0.02$ \cite{Farooq, Jesus, Garza}, while the
present value of the deceleration parameter is  $q_{0}=-0.50 \pm0.01$ \cite{Capozziello, Mamon, Basilakos},
indicating that the phase is accelerating.

\section{$Om(z)$ diagnostic and linear perturbations}
\label{sec5}

\subsection{$Om(z)$ diagnostic}

Sahni et al. \cite{Omz} introduced the $Om(z)$ diagnostic parameter as an
alternative to the statefinder parameter, which aids in distinguishing the
current matter density contrast $Om$ in various models more successfully.
This is also a geometrical diagnostic that is clearly dependent on red-shift (%
$z$) and the Hubble parameter ($H$). It is defined as follows: 
\begin{equation}
Om\left( z\right) =\frac{\left( \frac{H\left( z\right) }{H_{0}}\right) ^{2}-1%
}{\left( 1+z\right) ^{3}-1}.
\end{equation}

\begin{figure}[h]
\centerline{\includegraphics[scale=0.70]{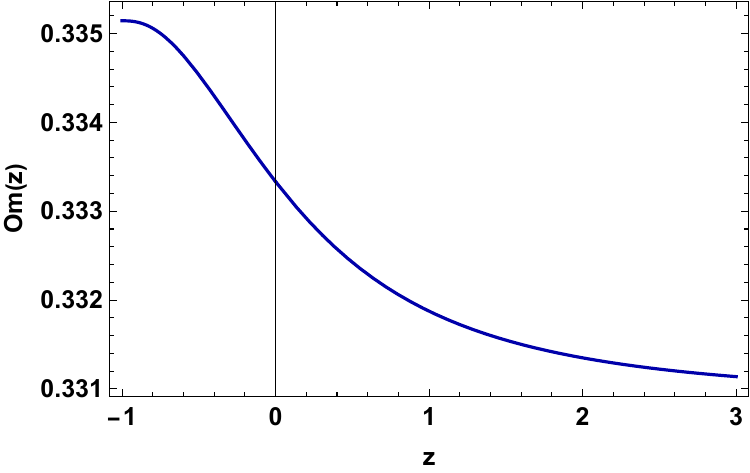}}
\caption{Evolution of the $Om(z)$ diagnostic parameter from the study of the
combined $OHD+BAO+SNe$ data for the best fitting values of $A$, $B$, and $m$.}
\label{Om}
\end{figure}

The negative slope of $Om(z)$ corresponds to quintessence type behavior $%
(-1<\omega< -1/3)$, while the positive slope corresponds to phantom-type
behavior $(\omega<-1)$. The $\Lambda$CDM model $(\omega=-1)$ is represented
by the constant nature of $Om(z)$. According to Fig. \ref{Om}, the $Om(z)$
diagnostic parameter has a negative slope throughout its entire domain. As a
result of the $Om(z)$ diagnostic test, our $f(Q)$ model follows the
quintessence scenario. Based on the findings, we can draw a conclusive inference that the behavior of the $Om(z)$ diagnostic parameter aligns with the behavior exhibited by the EoS parameter. The correspondence between these two parameters indicates a strong relationship, suggesting that variations in the EoS parameter are effectively captured by the $Om(z)$ diagnostic parameter. This observation underscores the utility and reliability of the $Om(z)$ parameter as a diagnostic tool for understanding the dynamics of the DE component.

\subsection{Linear perturbations}

In this subsection, our focus is on examining the stability of the $f(Q)$ cosmological model by analyzing the effects of linear homogeneity and isotropic perturbation. By considering small deviations from the Hubble parameter given by Eq. (\ref{H1}) and the energy density evolution i.e. Eq. (\ref{F1}), we aim to understand the behavior and robustness of the cosmological models under study. Linear perturbation analysis has been extensively used in cosmology to study the growth of structures and the evolution of the universe. Many previous studies have successfully employed linear approximations to explore the behavior of modified gravity theories and assess their compatibility with observational data \cite{Farrugia/2016,Dombriz/2012,Anagnost/2021,Mishra}. The perturbations under consideration in this analysis are of first order,
\begin{equation}
\widehat{H}(t)=H(t)(1+\delta \left( t\right) )  \label{Hp}
\end{equation}%
\begin{equation}
\widehat{\rho}(t)=\rho (t)(1+\delta _{m}\left( t\right) ),
\end{equation}%
where $\delta \left( t\right) $ represents the isotropic deviation of the background Hubble parameter, while $\delta_{m}\left( t\right) $ corresponds to the matter overdensity. Hence, the
perturbation of the functions $f(Q)$ and $f_{Q}$ can be expressed as $\delta
f=f_{Q}\delta Q$ and $\delta f_{Q}=f_{QQ}\delta Q$, where $\delta
Q=12H\delta H$ is the first-order perturbation of the scalar $Q$. So,
neglecting the higher power of $\delta \left( t\right) $, the Hubble
parameter can be expressed as $6\widehat{H}^{2}=6H^{2}\left( 1+\delta \left(
t\right) \right) ^{2}=6H^{2}\left( 1+2\delta \left( t\right) \right) $. Now,
using Eq. \eqref{F1} we get%
\begin{equation}
Q\left( f_{Q}+2Qf_{QQ}\right) \delta =-\rho \delta _{m}.
\end{equation}

This gives rise to the matter-geometric perturbation relation, and the
perturbed Hubble parameter can be calculated using Eq. (\ref{Hp}). Then, just use perturbation continuity equation to get the analytical solution to
the perturbation function,  
\begin{equation}
\dot{\delta _{m}}+3H(1+\omega )\delta =0.
\end{equation}

Solving the above equations for $\delta $ and $\delta _{m}$ yields the first
order differential equation, 
\begin{equation}
\dot{\delta _{m}}-\frac{3H(1+\omega )\rho }{Q(f_{Q}+2Qf_{QQ})}\delta _{m}=0.
\end{equation}

Using Eqs. (\ref{F1}) and (\ref{F2}) to simplify the previous equation once
more, the solution is expressed as, 
\begin{equation}
\delta _{m}\left( z\right)  =\delta _{m_{0}}H\left( z\right),
\label{dmm}
\end{equation}
and 
\begin{equation}
\delta \left( z\right)  =-\frac{\delta _{m0}}{3\left( 1+\omega
_{eff}\right) }\frac{\overset{.}{H}}{H}.
\label{dd}
\end{equation}

By using Eqs. (\ref{H1}) and (\ref{dmm}), we obtain
\begin{equation}
    \delta _{m}\left( z\right)  =\delta _{m0}H_{0}\left( \frac{A(z+1)^{3}+B}{A+B}\right) ^{\frac{1}{2m+2}}.
\end{equation}

Again, by using Eqs. (\ref{17}) and (\ref{dd}), we obtain

\begin{equation}
    \delta \left( z\right)  =\frac{\delta _{m0}H_{0}\left( \frac{A(z+1)^{3}+B}{A+B}\right) ^{\frac{1}{%
2m+2}}}{2(m+1)}.
\end{equation}

\begin{figure}[h]
\centerline{\includegraphics[scale=0.70]{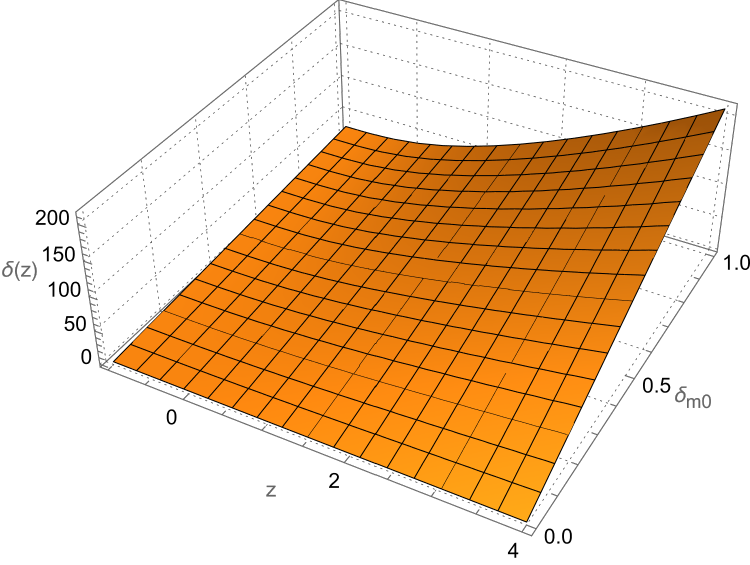}}
\caption{Evolution of the $\delta(z)$ from the study of the
combined $OHD+BAO+SNe$ data for the best fitting values of $A$, $B$, and $m$ and different $\delta_{m0}$ values.}
\label{d}
\end{figure}

\begin{figure}[h]
\centerline{\includegraphics[scale=0.70]{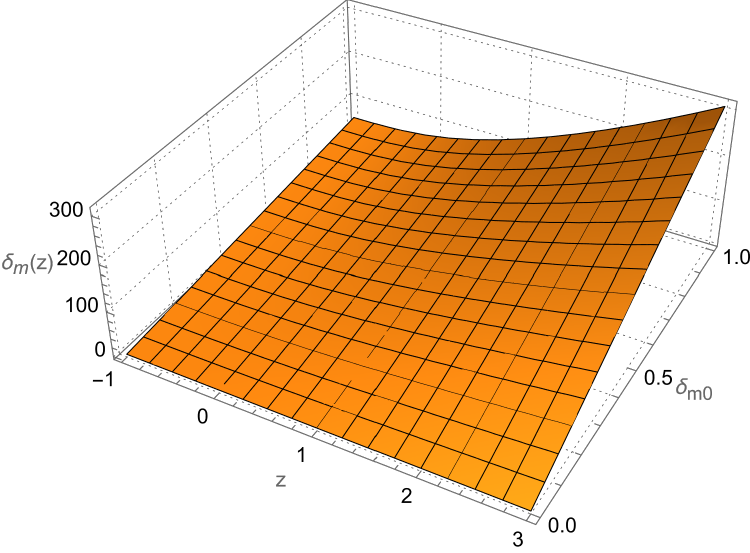}}
\caption{Evolution of the $\delta_{m}(z)$ from the study of the
combined $OHD+BAO+SNe$ data for the best fitting values of $A$, $B$, and $m$ and different $\delta_{m0}$ values.}
\label{dm}
\end{figure}

Figs. \ref{d} and \ref{dm} show the history of the perturbation terms $\delta
_{m}\left( z\right) $ and $\delta \left( z\right) $ in terms of red-shift $z$%
. Both the perturbations $\delta _{m}\left( z\right) $ and $\delta \left(
z\right) $ diminish rapidly and reach zero at late times. It may also be
demonstrated that the behavior of  $\delta _{m}\left( z\right) $ and $\delta
\left( z\right) $ is the same for all model parameter values. Consequently,
using the scalar perturbation approach, our $f(Q)$ model demonstrates stable
behavior.

\section{Conclusion}
\label{sec6}

The current scenario of accelerating Universe expansion is now a significant topic of study. Two approaches have been proposed to explain this cosmic acceleration. One approach is to investigate different dynamical DE models (such as quintessence and phantom), while another is to analyze alternate gravity theories. In this paper, we investigated accelerated expansion using the FLRW Universe and the $f(Q)$ theory of gravity, particularly $f(Q)=\beta Q^{\left( m+1\right) }$,
where $\beta$ and $m$ are arbitrary constants. We obtained the solution of the Hubble parameter using the parametrization form of the effective EoS parameter as $\omega _{eff}=-1+\frac{A}{A+B\left( 1+z\right) ^{-3}}$ (where $A$
and $B$ are arbitrary constants), which leads to a varying deceleration parameter. As shown in Sec. \ref{sec4} of this work, we constrained model parameters ($A$, $B$, and $m$) using the MCMC approach with a combined analysis of $OHD$, $BAO$, and $SNe$ data. The best-fit values obtained are $A=0.342^{+0.022}_{-0.022}$, $%
B=0.677^{+0.025}_{-0.025}$, and $m=0.013^{+0.021}_{-0.021}$. For $m=0$, the best-fit $A=0.3353\pm 0.0010$, and $B=0.6837 \pm0.0019$. Furthermore, with the constrained values of $A$, $B$, and $m$ from the
combined $OHD+BAO+SNe$ data, we analyzed the behavior of the density parameter, EoS parameter, and deceleration parameter as a function of red-shift, as shown in Figs. \ref{rho}, \ref{Omega}, and \ref{q}. Fig. \ref{rho} shows that both the matter density parameter and the DE density parameter are increasing functions of red-shift and exhibit the expected positive behavior. The evolution of the EoS parameter in Fig. \ref{Omega} supported the accelerating nature of the Universe's expansion phase, and the model behaves like a quintessence in the present. Furthermore, the present value of the EoS parameter for DE is estimated to be $\omega_{0}=-0.91 \pm 0.08$. Fig. \ref{q} indicates that the model transitions
from a decelerated stage to an accelerated stage. The present
value of the transition red-shift is $z_{tr}=0.60 \pm 0.02$ based on constrained values of model parameters $A$, $B$, and $m$ from the combined $OHD+BAO+SNe$ data, whereas the
present value of the deceleration parameter is $q_{0}=-0.50 \pm0.01$, showing that the phase is accelerating.

Then we evaluated the $Om(z)$ diagnostic parameter for our presumed $f(Q)$ model. As a result, we observed that the behavior of the $Om(z)$ diagnostic parameter conforms to the behavior of the effective EoS parameter. Lastly, the perturbation terms shown in Figs. \ref{d} and \ref{dm} confirmed that the model is stable under the scalar perturbation method. Based on our analysis, we reach a compelling conclusion that our $f(Q)$ cosmology, incorporating the effective EoS parameter form, offers a highly efficient framework for explaining various late-time cosmic phenomena in the Universe. The fact that this model demonstrates observational validity further supports its credibility and reliability. 
\section*{Acknowledgments}
This research is funded by the Science Committee of the Ministry of Science and Higher Education of the Republic of Kazakhstan (Grant No. AP09058240).

\textbf{Data availability} This article does not present any novel or additional data.\newline



\begin{thebibliography}{99}

\bibitem{Riess} A.G. Riess et al., \textit{Astron. J.} \textbf{116}, 1009
(1998).

\bibitem{Perlmutter} S. Perlmutter et al., \textit{Astrophys. J.} \textbf{517%
}, 565 (1999).

\bibitem{D.J.} D.J. Eisenstein et al., \textit{Astrophys. J.} \textbf{633},
560 (2005).

\bibitem{W.J.} W.J. Percival at el., \textit{Mon. Not. R. Astron. Soc.} 
\textbf{401}, 2148 (2010).

\bibitem{R.R.} R.R. Caldwell, M. Doran, \textit{Phys. Rev. D} \textbf{69},
103517 (2004).

\bibitem{Z.Y.} Z.Y. Huang et al., \textit{J. Cosm. Astrop. Phys.} \textbf{%
0605}, 013 (2006).

\bibitem{T.Koivisto} T. Koivisto, D.F. Mota, \textit{Phys. Rev. D} \textbf{73%
}, 083502 (2006).

\bibitem{S.F.} S.F. Daniel, \textit{Phys. Rev. D} \textbf{77}, 103513 (2008).

\bibitem{Capo/2008} S. Capozziello, V.F. Cardone, V. Salzano, Phys. Rev. D, \textbf{78}, 063504 (2008).

\bibitem{Nojiri/2007} S. Nojiri, S. D. Odintsov, Phys. Lett. B \textbf{657}, 238 (2007).

\bibitem{Harko/2011} T. Harko et al., Phys. Rev. D, \textbf{84}, 024020 (2011).

\bibitem{Momeni/2015} D. Momeni, R. Myrzakulov, E. Gudekli, Int. J. Geom. Meth. Mod.
Phys. \textbf{12}, 1550101 (2015).

\bibitem{Capo/2011} S. Capozziello et al., Phys. Rev. D, \textbf{84}, 043527 (2011).

\bibitem{Nunes/2016} R.C. Nunes, S. Pan, E.N. Saridakis, J. Cosm. Astropart. Phys., \textbf{08}, 011 (2016). 

\bibitem{Q0} J. B. Jimenez et al., \textit{Phys. Rev. D} \textbf{98}, 044048
(2018).

\bibitem{Q1} J.B. Jimenez et al., \textit{Phys. Rev. D} \textbf{101}, 103507
(2020).

\bibitem{Q2} W. Khyllep et al., \textit{Phys. Rev. D} 103, 103521 (2021).

\bibitem{Q3} N. Dimakis, A. Paliathanasis, and T. Christodoulakis, \textit{%
Class. Quantum Gravity} \textbf{38}, 22 (2021).

\bibitem{Q4} T. Harko et al., \textit{Phys. Rev. D} \textbf{98}, 084043
(2018).

\bibitem{Q5} F. D Ambrosio et al., \textit{Phys. Rev. D} \textbf{105}, 024042 (2022).

\bibitem{Q6} M. Hohmann, \textit{Phys. Rev. D} \textbf{104}, 124077 (2021).

\bibitem{Q7} F. K. Anagnostopoulos, S. Basilakos, and E. N.Saridakis, \textit{Phys. Lett. B} \textbf{822}, 136634 (2021).

\bibitem{Q8} M. Hohmann, \textit{Phys. Rev. D} \textbf{99}, 024009 (2009).

\bibitem{Q9} B. J. Barros et al., \textit{Phys.Dark Univ.} \textbf{30}, 100616 (2020).

\bibitem{Q10} I. Soudi et al., \textit{Phys. Lett. B} \textbf{100}, 044008 (2019).

\bibitem{ko1} M. Koussour et al., \textit{Phys. Dark Univ.} \textbf{36}, 101051 (2022).

\bibitem{ko2} M. Koussour et al., \textit{J. High Energy Phys.} \textbf{37}, 15-24 (2023).

\bibitem{ko3} M. Koussour and M. Bennai, \textit{Chin. J. Phys.} \textbf{79}, 339-347 (2022).

\bibitem{ko4} M. Koussour et al.,{\ }\textit{Ann. Phys.} \textbf{445}, 169092 (2022).

\bibitem{ko5} M. Koussour et al., \textit{J. High Energy Astrophys, } \textbf{35}, 43-51 (2022).

\bibitem{OC1} R. Lazkoz et al., \textit{Phys. Rev. D} \textbf{100}, 104027 (2019).

\bibitem{OC2} I. Ayuso, R. Lazkoz, and V. Salzano, \textit{Phys. Rev. D} \textbf{103}, 063505 (2021).

\bibitem{OC3} S. A. Narawade and B. Mishra, \textit{arXiv preprint} arXiv:2211.09701 (2022).


\bibitem{P0} A. A. Mamon, \textit{Int. J. Mod. Phys. D} \textbf{26}, 1750136
(2017).

\bibitem{P1} Y. g. Gong and Y. Z. Zhang, \textit{Phys. Rev. D} \textbf{72},
043518 (2005).

\bibitem{P2} M. Chevallier and D. Polarski, \textit{Int. J. Mod. Phys. D} 
\textbf{10}, 213 (2001).

\bibitem{P3} E. V. Linder, \textit{Phys. Rev. Lett.} \textbf{90}, 091301
(2003).

\bibitem{P4} A. R. Cooray and D. Huterer, \textit{Astrophys. J.} \textbf{513}%
, L95 (1999).

\bibitem{P5} P. Astier, \textit{Phys. Lett. B} \textbf{500}, 8 (2001).

\bibitem{P6} J. Weller and A. Albrecht, \textit{Phys. Rev. D} \textbf{65},
103512 (2002).

\bibitem{P7} G. Efstathiou, \textit{Mon. Not. Roy. Astron. Soc.} \textbf{310}%
, 842 (1999).

\bibitem{P8} H. K. Jassal, J. S. Bagla and T. Padmanabhan, \textit{Phys.
Rev. D} \textbf{72}, 103503 (2005).

\bibitem{P9} E. M. Barboza, Jr. and J. S. Alcaniz, \textit{Phys. Lett. B} 
\textbf{666}, 415 (2008).

\bibitem{P10} ] E. V. Linder and D. Huterer, \textit{Phys. Rev. D} \textbf{72%
}, 043509 (2005).

\bibitem{P11} A. De Felice, S. Nesseris and S. Tsujikawa, \textit{JCAP} 
\textbf{1205}, 029 (2012).

\bibitem{P12} R. J. F. Marcondes and S. Pan, \textit{arXiv preprint arXiv}%
:1711.06157 (2017).

\bibitem{PW1} M. Koussour et al., \textit{Nucl. Phys. B.} 
\textbf{990}, 116158 (2023).

\bibitem{PW2} M. Koussour and A. De, \textit{Eur. Phys. J. C} 
\textbf{83}, 400 (2023).


\bibitem{Yu/2018} Yu, B. Ratra, F-Yin Wang, Astrophys. J., \textbf{856}, 3
(2018).

\bibitem{Moresco/2015} M. Moresco, Month. Not. R. Astron. Soc., \textbf{450}%
, , L16-L20 (2015).

\bibitem{Sharov/2018} G.S. Sharov, V.O. Vasilie, Mathematical Modelling and
Ge-ometry \textbf{6}, 1 (2018).

\bibitem{Blake/2011} C. Blake et al., Month. Not. R. Astron. Soc., \textbf{%
418}, 1707 (2011).

\bibitem{Percival/2010} W. J. Percival et al., Month. Not. R. Astron. Soc., 
\textbf{401}, 2148 (2010).

\bibitem{Giostri/2012} R. Giostri et al., J. Cosm. Astropart. Phys. \textbf{%
1203}, 027 (2012).

\bibitem{Scolnic/2018} D.M. Scolnic et al., Astrophys. J, \textbf{859}, 101
(2018).

\bibitem{Chang/2019} Z. Chang et al., Chin. Phys. C, \textbf{43}, 125102
(2019).

\bibitem{Mackey/2013} D. F. Mackey et al., \textit{Publ. Astron. Soc. Pac.} 
\textbf{125}, 306 (2013).

\bibitem{Kessler/2017} R. Kessler, D. Scolnic, Astrophys. J., \textbf{836},
56 (2017).

\bibitem{Planck2020} Planck Collaboration, Astron. Astrophys., \textbf{641},
A6 (2020).


\bibitem{Hernandez} A. Hernandez-Almada, et al.,\textit{\ Eur. Phys. J. C} 
\textbf{79}, 12 (2019).

\bibitem{Gruber} C. Gruber, O. Luongo,\textit{Phys. Rev. D} \textbf{89},
103506 (2014).

\bibitem{Farooq} O. Farooq, et al.,\textit{Astrophys. J.} \textbf{835},
26-37 (2017).

\bibitem{Jesus} J.F. Jesus, et al.,\textit{J. Cosmol. Astropart. Phys.} 
\textbf{04}, 053-070 (2020).

\bibitem{Garza} J.R. Garza, et al.,\textit{Eur. Phys. J. C} \textbf{79}, 890
(2019).

\bibitem{Capozziello} S. Capozziello, R. D Agostino and O. Luongo,\textit{%
Mon. Not. Roy. Astron. Soc.} \textbf{494}, 2576 (2020).

\bibitem{Mamon} S. A. Al Mamon and S. Das,\textit{\ Eur. Phys. J. C} \textbf{%
77}, 495 (2017).

\bibitem{Basilakos} S. Basilakos, F. Bauera and J. Sola,\textit{J. Cosmol.
Astropart. Phys.} \textbf{01}, 050-079 (2012).

\bibitem{Omz} V. Sahni, A. Shafieloo, and A. A. Starobinsky, \textit{Phys.
Rev. D} \textbf{78}, 103502 (2008).

\bibitem{Farrugia/2016} G. Farrugia, J. L. Said, Phys. Rev. D, \textbf{94},
124054 (2016).

\bibitem{Dombriz/2012} A. de la C-Dombriz, D. S-Gomez, Class. Quantum Grav., 
\textbf{29}, 245014 (2012).

\bibitem{Anagnost/2021} F. K. Anagnostopoulos, S. Basilakos, E. N.
Saridakis, Phys. Lett. B, \textbf{822}, 136634 (2021).

\bibitem{Mishra} S. A. Narawade et al., \textit{Phys. Dark Universe} \textbf{36}, 101020 (2022).

\end{thebibliography}
\end{document}